\documentstyle[preprint,aps,pre,graphics,epsfig,eqsecnum]{revtex}

\begin{document}
\draft
\title{{\bf Random Resistor-Diode Networks and the Crossover from Isotropic to
Directed Percolation}}
\author{{\sc Hans-Karl\ Janssen and Olaf\ Stenull}}
\address{Institut f\"{u}r Theoretische Physik III, Heinrich-Heine-Universit\"{a}t,\\
40225 D\"{u}sseldorf, Germany }
\date{\today}
\maketitle

\begin{abstract}
By employing the methods of renormalized field theory we show that the percolation
behavior of random resistor-diode networks near the multicritical line belongs
to the universality class of isotropic percolation. We construct a mesoscopic model  from the general epidemic process by including a relevant isotropy-breaking perturbation. We present a two-loop calculation of the crossover exponent $\phi$. Upon blending the $\varepsilon$-expansion result with the exact value $\phi =1$ for one dimension by a rational approximation, we obtain for two dimensions $\phi = 1.29\pm 0.05$. This value is in agreement with the recent simulations of a two-dimensional random diode network by Inui, Kakuno, Tretyakov, Komatsu, and Kameoka, who found an order parameter exponent $\beta$ different from those of isotropic and directed percolation. Furthermore, we reconsider the theory of the full crossover from isotropic to directed percolation by Frey, T\"{a}uber, and Schwabl and clear up some minor shortcomings.
\end{abstract}



\tightenlines

\section{Introduction}

Random resistor-diode networks (RDN) were introduced by Redner~\cite{Red81,Red82a,Red82b}. Nevertheless, they were already contained implicitly in the pioneering work of Broadbent and Hammersley~\cite{BrHa57} on percolation. RDN define a percolation model (for a recent introduction to percolation see Stauffer and Aharony~\cite{StaAh94}) on a $d$-dimensional hypercubic lattice in which nearest-neighbor sites are connected by a resistor, a positive diode (conducting only in a preferred direction), a negativ diode (conducting only opposite to the preferred direction), or an insulator with respective probabilities $p$, $p_{+}$, $
p_{-}$, and $q=1-p-p_{+}-p_{-}$. In the three dimensional phase diagram
(pictured as a tetrahedron spanned by the four probabilities) one finds a
nonpercolating and three percolating phases. The percolating phases are isotropic, positively directed, or negatively directed. Between the phases there are surfaces of continuous transitions. All four phases meet along a multicritical line,
where $0\leq r:=p_{+}=p_{-}\leq 1/2$ and $p=p_{c}(r)$. On the entire multicritical line, i.e., independently of $r$, one finds the scaling properties of usual isotropic percolation ($r=0$). 

About 20 years ago Redner~\cite{Red81,Red82a,Red82b} studied the
phase diagram sketched in Fig.~1 as well as geometrical properties of RDN in two dimensions. He used real-space renormalization methods and planar
lattice duality~\cite{DBP81}.

Recently Inui {\em et al}.~\cite{IKTK99} have measured the order parameter exponent $\beta $ for the special case of a two-dimensional random diode network (with $p=q=0$) by Monte Carlo methods combined with series expansions. At
the symmetric critical point $p_{+}=p_{-}=1/2$ (so called random Manhattan
(RM)), they found $\beta =0.1794\pm 0.008$, which does not coincide with the
known values wether for isotropic percolation (IP), $\beta_{IP}=5/36$, nor directed  percolation (DP), $\beta_{DP}=0.27643$. Therefore, they concluded
that the percolation properties of the random diode network constitute a
new universality class different from isotropic and directed percolation.

In this paper we study RDN by the methods of renormalized field theory. Contrary to 
Inui {\em et al}.\ we find that random diode networks at the percolation point as well as RDN at the full multicritical line belong to the universality class of isotropic percolation. The variable $r=p_{+}=p_{-}$, which
maps out the multicritical line $p=p_{c}(r)$, is a redundant
variable in the sense of the classification scheme of scaling variables by
Wegner \cite{We74}. Thus, all the points of this line are equivalent to the
usual isotropic percolation point with $r=0$.

The case $r>0$, in which positive and negative diodes are distributed with equal probability, leads to a breaking of isotropy and to elongated percolating clusters. However, this symmetry breaking can be easily compensated in the mesoscopic field theoretic formulation by a simple rescaling of the length scale of the preferred direction. The specific length scale is therefore
redundant. Wegner showed that the renormalization flow of a redundant
variable depends on the particular form of renormalization group used and does not
affect the physics. Thus, in the case of Redner's real space renormalization
group, a special fixed point, the so called mixed one, is distinguished.

The situation is different for a symmetry breaking which favors not only an
axis but also a direction on that axis. This leads to a relevant variable $\sim
(p_{+}-p_{-})$ and therefore to a new correlation length exponent $\nu =\phi
\nu _{IP}$. The crossover exponent $\phi $ describes the beginning of
the crossover to directed percolation. It also relates the order parameter
exponents, $\beta =\phi \beta _{IP}$. In two dimensions, it has, according
to the results of Inui {\em et al}.~\cite{IKTK99}, the value $\phi =1.29...$~.
Our perturbation calculation yields $\phi =1.29\pm0.05$ in agreement with the simulations. It should be noted that the interpretation of the multicritical line as a line of equivalent fixed points with the same scaling behavior, as well as the new correlation exponent $\nu$, can be already found implicitly and explicitly in the early papers of Redner \cite{Red81,Red82a,Red82b} (his notation is $\nu _{+}$ for
the new exponent $\nu$).

The organization of the remainder of this paper is as follows. In Sec.~II,
we first develop a mesoscopic field theoretic model that is capable to
describe the crossover from isotropic to directed percolation, which is the
basic feature of RDN. We describe the renormalization of the model and
calculate the renormalization factors to two-loop order. In Sec.~III, we
introduce the renormalization group equation for the model and derive the
general asymptotic scaling properties. In Sec.~IV, we derive an
interpolating formula for the crossover exponent from the two-loop $\varepsilon $-expansion and show that it reproduces the result of Inui {\em et al}. In Sec.~V we reconsider the theory by Frey, T\"{a}uber, and Schwabl~\cite{FTS94} for the crossover from isotropic to directed percolation and clear up some minor shortcomings. In Sec.~VI, we give some conclusions and summarize our work. In an appendix, we present the two-loop calculation of the renormalization factors.

\section{The Field Theoretic Model and its Renormalization}

Here we develop a mesoscopic model that is capable to describe the crossover from isotropic to directed percolation. In this paper we are only interested in connectivity properties of the percolating system. In contrast to earlier work on random resistor networks \cite{stenull_janssen_oerding_99,janssen_stenull_oerding_99,janssen_stenull_99,stenull_janssen_00} we neglect transport properties as the conductance etc. 
In other words: all we ask is wether two points on the lattice are connected or not. Formally, we consider the limit of zero resistance of the conducting elements.

Percolating clusters in space and time can be generated by a stochastic spreading process known as the general epidemic process (GEP) \cite{Gra83}. In order to apply field-theoretic methods \cite{Am84,ZiJu93}, it is convenient to use the path-integral representation of the underlying stochastic process $s({\bf x},t)$ \cite{Ja76,DeDo76,Ja92}. With the imaginary-valued response field denoted by $
\widetilde{s}({\bf x},t)$, the generating functional of the Greens
functions, the connected response, and correlation functions takes the form
\begin{equation}
{\cal W}\bigl[H,\widetilde{H}\bigr]=\ln \int {\cal D}\bigl(\widetilde{s},s
\bigr)\exp \biggl[-{\cal J}\bigl[\widetilde{s},s\bigr]+\int d^{d}x\int dt
\bigl(Hs+\widetilde{H}\widetilde{s}\bigr)\biggr]\ .  \label{PathInt}
\end{equation}
The dynamic functional ${\cal J}\bigl[\widetilde{s},s\bigr]$ and the
functional measure ${\cal D}\bigl(\widetilde{s},s\bigr)$ (${\cal D}\bigl(\widetilde{s},s\bigr)$ is a symbolic notation for $\prod_{{\bf x},t} \bigl( d\widetilde{s}({\bf x} ,t) ds({\bf x},t) \bigr) $ times a constant) are understood to be defined using a prepoint (Ito) discretization with respect to time \cite{Ja92}. The prepoint discretization leads to the causality rule $\theta (t\leq 0)=0$ in response functions. This causality rule then forbids response propagator loops in the diagrammatical perturbation expansion.

Using the path-integral formulation a renormalized field-theory of dynamic isotropic percolation can be gained from the GEP~\cite{Ja85,CaGra85}. If we add to this model a relevant coupling which breaks isotropy and introduce a preferred direction ${\bf n}$ for the spreading of the disease in the $d$-dimensional space, we get the dynamic functional
\begin{equation}
{\cal J}=\int d^{d}x\Biggl\{\int dt\widetilde{s}\biggl[\partial _{t}+\lambda
\Bigl(\tau -\nabla ^{2}+2v\bigl({\bf n}\cdot \nabla \bigr)\Bigr)-\frac{%
\lambda g}{2}\widetilde{s}\biggr]s+\frac{\lambda ^{2}g}{2}\widetilde{s}%
\partial _{t}\biggl[\int^{t}dt^{\prime }s(t^{\prime })\biggr]^{2}\Biggr\}\ .
\label{JFunkt}
\end{equation}
This functional corresponds to the Langevin equations 
\begin{eqnarray}
\partial _{t}s({\bf x},t) &=&\lambda \Bigl(\nabla ^{2}-2v\bigl({\bf n}\cdot
\nabla \bigr)-\tau -gn({\bf x},t)\Bigr)s({\bf x},t)+\zeta ({\bf x},t)\ ,
\label{LangEq} \\
n({\bf x},t) &=&\lambda \int_{-\infty }^{t}s({\bf x},t^{\prime })dt^{\prime
}\ , \\
\langle \zeta ({\bf x},t)\zeta ({\bf x}^{\prime },t^{\prime })\rangle
&=&\lambda gs({\bf x},t)\delta ({\bf x-x}^{\prime })\delta (t-t^{\prime }) \ ,
\end{eqnarray}
for the GEP with the suitable scaled density $s({\bf x},t)$ of the infected individuals. Here $n({\bf x},t)$ constitutes the density of the immun (or dead)
individuals and $\zeta ({\bf x},t)$ is a Gaussian noise which is zero in
spatial regions where the disease is extinguished. The deterministic drift
of the disease in space is represented by the flow 2$\lambda v{\bf n}s({\bf x%
},t)$.

Of course, if the $d$-dimensional rotational symmetry is broken to $(d-1)$
-dimensional isotropy, the diffusion constants for longitudinal (with
respect to the direction ${\bf n}$) and transversal spreading are in general
different. Thus we have to consider a more general diffusion operator $
\lambda \nabla ^{2}$ $\rightarrow \lambda (\nabla _{\bot
}^{2}+c^{-2}\partial _{\Vert }^{2})$. However, it is easy to see that the
new parameter $c$ can be absorbed into the definition of the longitudinal
length scale by $cx_{\Vert }\rightarrow x_{\Vert }$ followed by an
appropriate change of the densities and the coupling constant $g$. After that,
diffusional spreading looks isotropic again. The parameter $c$ depends
on the microscopic model. It is a redundant variable in
the sense of Wegner~\cite{We74} and is responsible for the multicritical
line in the RDN~\cite{Red81,Red82a,Red82b}. From the microscopic
RDN-standpoint, the three variables $\tau $, $c$, $v$ are analytical
functions of the three probabilities $p$, $p_{+}$, $p_{-}$ for the resistors
and diodes near the critical manifolds of RDN and share their spatial
symmetries. Thus one has $\tau (p,p_{+},p_{-})=\tau (p,p_{-},p_{+})$, $
c(p,p_{+},p_{-})=c(p,p_{-},p_{+})$, and $v(p,p_{+},p_{-})=-v(p,p_{-},p_{+})$. In particular we have $c(p,0,0)=1$, but in general $c(p,r,r)\neq 1$ if $r>0$.
Moreover, $v(p,r,r)=0$ holds. Remember that $p_{+}=p_{-}=r$ and $p=p_{c}(r)$
defines the multicritical line.

In this paper we are interested only in the static behavior of the process, i.e., in the statistics of the distributions of immunes in space $n(%
{\bf x},\infty )$ after a long time when the epidemic is extinguished. These
distributions constitute isotropic percolating clusters at the critical
point of the GEP, which is given by $v=v(p_{c}(r),r,r)=0$ and $\tau =\tau
(p_{c}(r),r,r)=\tau _{c}=0$ if we neglect fluctuation corrections. The
density $n({\bf x},\infty )$ is proportional to the Fourier transform of $s$
for frequency zero. The statistical weight for these frequency-zero modes
can be found by invoking the formal limit
\begin{equation}
\widetilde{s}({\bf x},t)\rightarrow \widetilde{\varphi }({\bf x})=\text{%
const.\ },\qquad \lambda \int_{-\infty }^{\infty }dts({\bf x,}t){\bf =}n(%
{\bf x},\infty )\rightarrow \varphi ({\bf x)}\ ,  \label{StatFeld}
\end{equation}
for the fields in the dynamic functional ${\cal J}$, Eq.\ (\ref{JFunkt}).
This manipulation can be controlled term by term in a diagrammatical
perturbation expansion and leads to the wanted representation of the zero
frequency Green's functions as functional integrals with a weight $\exp (-
{\cal H})$ and a quasistatic Hamiltonian
\begin{equation}
{\cal H}=\int d^{d}x\Biggl\{\widetilde{\varphi }\biggl[\tau -\nabla ^{2}+2v
\bigl({\bf n}\cdot \nabla \bigr)+\frac{g}{2}\Bigl(\varphi -\widetilde{
\varphi }\Bigr)\biggr]\varphi \Biggr\}\ .  \label{Hamilt}
\end{equation}
Remember that closed loops of the response propagators are forbidden in the Feynman diagrams as a consequence of the causality rule. Therefore, the functional ${\cal H}$ is, up to a rescaling, identical with the statistical functional
considered by Frey, T\"{a}uber, and Schwabl~\cite{FTS94,BeCa84} in their
work on the crossover from isotropic to directed percolation.

To absorb ultraviolet divergencies in a perturbational calculation of the
Green's functions with the Hamiltonian ${\cal H}$  we
use, in the case $v=0$, the following renormalization scheme~\cite{Ja85}:
\begin{eqnarray}
\widetilde{\varphi } &\rightarrow &\mathaccent"7017{\widetilde{\varphi }}
=Z^{1/2}\widetilde{\varphi }\ ,\quad \varphi \rightarrow \mathaccent"7017{
\varphi }=Z^{1/2}\varphi \ ,  \label{Ren1a} \\
\tau &\rightarrow &\mathaccent"7017{\tau }=Z^{-1}Z_{\tau }\tau \mu ^{2}+
\mathaccent"7017\tau _{c}\ ,\quad g^{2}\rightarrow \mathaccent"7017{g}
^{2}=G_{\varepsilon }^{\,-1}Z^{-3}Z_{u}u\mu ^{\varepsilon }\ .  \label{Ren1b}
\end{eqnarray}
Here $\varepsilon =6-d$, $G_{\varepsilon }=\Gamma (1+\varepsilon /2)/(4\pi
)^{d/2}$, and $\mu $ is the usual external momentum scale, which makes the
renormalized coupling constant $u$ dimensionless. Note that the fields $
\widetilde{\varphi }$ and $\varphi $ are renormalized by the same $Z$-factor
as a consequence of the reflection symmetry $\varphi ({\bf x}_{\bot
},x_{\Vert })\leftrightarrow -\widetilde{\varphi }({\bf x}_{\bot },-x_{\Vert
})$ of ${\cal H}$, which leads eventually to the equality $\beta ^{\prime
}=\beta $ between the exponents characterizing the particle density and the
percolation probability. The renormalizations are known from percolation
field theory up to three-loop order \cite{AKM81}. Using dimensional
regularization and minimal subtraction (minimal renormalization) together with the $\varepsilon $-expansion one finds $\mathaccent"7017\tau _{c}=0$ and to two-loop order
\begin{eqnarray}
Z &=&1+\frac{u}{6\varepsilon }+\biggl(11-\frac{37}{12}\varepsilon \biggr)
\biggl(\frac{u}{6\varepsilon }\biggr)^{2}+O\bigl(u^{3}\bigr)\ ,  \label{Z} \\
Z_{\tau } &=&1+\frac{u}{\varepsilon }+\biggl(9-\frac{47}{12}\varepsilon 
\biggr)\biggl(\frac{u}{2\varepsilon }\biggr)^{2}+O\bigl(u^{3}\bigr)\ ,
\label{Zt} \\
Z_{u} &=&1+\frac{4u}{\varepsilon }+\biggl(15-\frac{59}{12}\varepsilon \biggr)
\biggl(\frac{u}{\varepsilon }\biggr)^{2}+O\bigl(u^{3}\bigr)\ .  \label{Zu}
\end{eqnarray}

If $v\neq 0$, further renormalizations are needed. Because $v\sim \mu $,
this relevant parameter has positive naive dimension like $\tau $. Hence, we
consider it as a soft variable (which means that the renormalization
constants do not depend on $\tau $ and $v$) as long as $v$ is finite and
complete the renormalization scheme by
\begin{equation}
\tau \rightarrow \mathaccent"7017{\tau }=Z^{-1}\biggl(Z_{\tau }\tau
+Y_{v\tau }v^{2}\biggr)\mu ^{2}\ ,\quad v\rightarrow \mathaccent"7017{v}
=Z^{-1}Z_{v}v\mu \ .  \label{Ren1c}
\end{equation}
Here we anticipate that $\tau $ and $v^{2}$ get mixed under renormalization.

To calculate $Z_{v}$ und $Y_{v\tau }$, we need the (unrenormalized)
propagator $G({\bf q})=\langle \varphi _{{\bf q}}\widetilde{\varphi }_{-{\bf 
q}}\rangle _{0}^{(trunc)}$. We define the spatial Fourier transform by $
\varphi ({\bf x})=\int_{{\bf q}}\varphi _{{\bf q}}\exp (i{\bf q\cdot x})$
(with the abbreviation $\int_{{\bf q}}=(2\pi )^{-d}\int d^{d}q$) and
get
\begin{equation}
G({\bf q})=\frac{1}{\tau +q^{2}+2i{\bf v}\cdot {\bf q}}=\frac{1}{\bigl(\tau
+v^{2}\bigr)+\bigl({\bf q}+i{\bf v}\bigr)^{2}}\ ,  \label{Prop}
\end{equation}
where we used the notation ${\bf v=}v{\bf n}$. The one-loop contribution to
the vertex function $\Gamma _{1,1}$ (the amputated one-particle irreducible
Green's function with one $s$- and one $\widetilde{s}$-leg) is given by
\begin{eqnarray}
\Sigma ^{(1)}({\bf q}) &=&\frac{g^{2}}{2}\int_{{\bf k}}G({\bf k+q}/2)G(-{\bf 
k}+{\bf q/}2)  \nonumber \\
&=&\frac{g^{2}}{2}\int_{0}^{\infty }ds_{1}ds_{2}\int_{{\bf k}}\exp \biggl[-
\bigl(s_{1}+s_{2}\bigr)\bigl(\tau +v^{2}\bigr)  \nonumber \\
&&\qquad \qquad \qquad \qquad \qquad -s_{1}\bigl({\bf k}+({\bf q}/2+i{\bf v)}
\bigr)^{2}-s_{2}\bigl({\bf k}-({\bf q}/2+i{\bf v)}\bigr)^{2}\biggr]
\nonumber \\
&=&-\frac{2G_{\varepsilon }g^{2}}{(2-\varepsilon )\varepsilon }\bigl(\tau +Q
\bigr)^{1-\varepsilon /2}K_{\varepsilon -4}^{(0)}\biggl(\sqrt{\frac{v^{2}-Q}{
\tau +Q}}\biggr)  \nonumber \\
&=&-G_{\varepsilon }g^{2}\tau ^{-\varepsilon /2}\biggl(\frac{2\tau }{
(2-\varepsilon )\varepsilon }K_{\varepsilon -4}^{(0)}\bigl(v/\sqrt{\tau }
\bigr)+\frac{Q}{\varepsilon }K_{\varepsilon -2}^{(1)}\bigl(v\sqrt{\tau }
\bigr)  \nonumber \\
&&\qquad \qquad \qquad \qquad \qquad -\frac{Q^{2}}{4}K_{\varepsilon }^{(2)}
\bigl(v/\sqrt{\tau }\bigr)+O(Q^{3})\biggr)\ ,  \label{SelfEn}
\end{eqnarray}
where we have defined the functions
\begin{equation}
K_{\alpha }^{(n)}(p)=\int_{0}^{1}dx\frac{\bigl(1-x^{2}\bigr)^{n}}{\bigl(
1+p^{2}x^{2}\bigr)^{1+\alpha /2}}  \label{K-Funkt}
\end{equation}
and the abbreviation
\begin{equation}
Q=q^{2}/4+i{\bf v\cdot q\ .}
\end{equation}
The expansion of the yet unrenormalized one-loop selfenergy $\Sigma ^{(1)}(
{\bf q})$ in $v^{2}$ and $\varepsilon $ yields
\begin{equation}
\Sigma ^{(1)}({\bf q})=-\frac{G_{\varepsilon }g^{2}\tau ^{-\varepsilon /2}}{
\varepsilon }\biggl(\Bigl(1+\frac{\varepsilon }{2}\Bigr)\tau +\frac{v^{2}}{3}
+\frac{2i}{3}{\bf v}\cdot {\bf q} + \frac{1}{6}{\bf q}^{2}+O(\varepsilon ^{2})\biggr)
\ .  \label{SelfEn2}
\end{equation}
Using the renormalization scheme Eqs.~(\ref{Ren1a},\ref{Ren1b},\ref{Ren1c}
), the renormalized vertex function $\Gamma _{1,1}$ is found to first order
in the renormalized coupling constant $u$ as
\begin{equation}
\Gamma _{1,1}=\biggl(Z_{\tau }\tau +Y_{v\tau }v^{2}-\frac{u}{\varepsilon }
\Bigl(\tau +\frac{v^{2}}{3}\Bigr)\biggr)\mu ^{2}+2i\biggl(Z_{v}-\frac{u}{
3\varepsilon }\biggr)\mu {\bf v}\cdot {\bf q} + \biggl(Z-\frac{u}{6\varepsilon
}\biggr) {\bf q}^{2}+O(\varepsilon ^{0})\ ,  \label{Vert11}
\end{equation}
from which the new renormalizations can be gathered to $O(u)$ as
\begin{eqnarray}
Z_{v} &=&1+\frac{u}{3\varepsilon }+\biggl(23-\frac{73}{12}\varepsilon \biggr)
\biggl(\frac{u}{6\varepsilon }\biggr)^{2}+O\bigl(u^{3}\bigr)\ ,  \label{Zv}
\\
Y_{v\tau } &=&\frac{u}{3\varepsilon }+\biggl(30-\frac{35}{2}\varepsilon 
\biggr)\biggl(\frac{u}{6\varepsilon }\biggr)^{2}+O\bigl(u^{3}\bigr) \ .
\label{Yvt}
\end{eqnarray}
Here we have included also the two-loop result calculated in the appendix.

\section{Renormalization Group Equation and Scaling Behavior}

Next we explore the scaling properties of percolation in the RDN
system. Scaling properties describe how physical quantities will transform
under a change of length scales. By the renormalization, we have introduced
the arbitrary mesoscopic length scale $\mu^{-1} $. The freedom to choose $
\mu $, keeping the unrenormalized fields and bare parameters $\left\{ 
\mathaccent"7017{\tau },\mathaccent"7017{v},\mathaccent"7017{g}\right\} $,
and, in cutoff regularization, the momentum cutoff $\Lambda $ fixed, can be
used to derive in a routine fashion the renormalization group (RG) equation
for the connected Green's functions
\begin{equation}
G_{N,\widetilde{N}}(\{{\bf x}\})=\biggl\langle\prod_{i=1}^{N}\varphi ({\bf x}
_{i})\prod_{j=N+1}^{N+\widetilde{N}}\widetilde{\varphi }({\bf x}_{j})
\biggr\rangle^{(conn)}\ .
\end{equation}
We denote $\mu$-derivatives at fixed bare parameters by $\left. \partial_{\mu } \right|_{0}$. From $\mu \left. \partial_{\mu} \right|_{0}
{\mathaccent"7017 G}_{N,\widetilde{N}}=0$ and the renormalization scheme, Eqs.~(\ref{Ren1a},\ref{Ren1b},\ref{Ren1c}), which lead to ${ \mathaccent"7017 G}_{N, 
\widetilde{N}}=Z^{(N+\widetilde{N})/2} \, G_{N,\widetilde{N}}$, we then find
the renormalization group (RG) equations
\begin{equation}
\biggl[{\cal D}_{\mu }+\frac{N+\widetilde{N}}{2}\gamma \biggr]G_{N,%
\widetilde{N}}=0  \ . \label{RGG}
\end{equation}
${\cal D}_{\mu }$ stands for the renormalization group differential operator
\begin{equation}
{\cal D}_{\mu }=\mu \partial _{\mu }+\beta _{u}\partial _{u}+\Bigl(\tau
(\kappa _{\tau }-2)+v^{2}\kappa _{v\tau }\Bigr)\partial _{\tau }+v(\kappa
_{v}-1)\partial _{v}\ .  \label{RGOp}
\end{equation}
Here we have introduced the Gell-Mann-Low functions 
\begin{eqnarray}
\beta _{u} &=&\left. \frac{\partial u}{\partial \ln \mu }\right| _{0}=\biggl(%
-\varepsilon +3\gamma -\gamma _{u}\biggr)u\ ,  \label{betau} \\
\tau \kappa _{\tau }+v^{2}\kappa _{v\tau } &=&\left. \frac{\partial \tau }{%
\partial \ln \mu }\right| _{0}=\tau \biggl(\gamma -\gamma _{\tau }\biggr)%
-v^{2}\gamma _{v\tau }\ ,  \label{kappat} \\
v\kappa _{v} &=&\left. \frac{\partial v}{\partial \ln \mu }\right| _{0}=v%
\biggl(\gamma -\gamma _{v}\biggr)\ ,  \label{kappav}
\end{eqnarray}
and the Wilson-functions $\gamma _{\cdots }=\partial \ln Z_{\cdots
}/\partial \ln \mu |_{0}$.

The RG equations can be solved in terms of a single flow parameter $l$ using the
characteristics
\begin{eqnarray}
l\frac{d}{dl}\bar{u}(l) &=&\beta _{u}(\bar{u}(l))\ ,\qquad \bar{u}(1)=u\ ,
\label{uflow} \\
l\frac{d}{dl}\bar{v}(l) &=&\bar{v}(l)\bigl(\kappa _{v}(\bar{u}(l))-1\bigr)\
,\qquad \bar{v}(1)=v\ ,  \label{vflow} \\
l\frac{d}{dl}\bar{\tau}(l) &=&\Bigl(\bar{\tau}\bigl(\kappa _{\tau }(\bar{u}%
(l))-2\bigr)+\bar{v}^{2}\kappa _{v\tau }(\bar{u}(l))\Bigr)\ ,\qquad \bar{\tau%
}(1)=\tau \ .  \label{tflow}
\end{eqnarray}
With help of these flow equations we recast Eq.~(\ref{RGG}) as
\begin{equation}
\biggl[l\frac{d}{dl}+\frac{N+\widetilde{N}}{2}\gamma (\bar{u}(l))\biggr]G_{N,%
\widetilde{N}}(\{{\bf x}\},\bar{\tau}(l),\bar{v}(l),\bar{u}(l)\},l\mu )=0\ .
\label{Gflow}
\end{equation}
Equations~(\ref{uflow}-\ref{tflow}) describe how the
parameters transform if we change the momentum scale $\mu $ according to $%
\mu \rightarrow \bar{\mu}(l)=l\mu $. Being interested in the infrared (IR)
behavior of the theory, we must study the limit $l\rightarrow 0$. According to Eq.~(\ref{uflow}) we expect that in this IR limit the coupling constant $\bar{u}(l)$ flows to a stable fixed point $u_{\ast }$ with $\beta _{u\ast }=0$. At the fixed point it is legitimate to diagonalize the part
of the RG differential operator containing the relevant parameters $\tau $ and $v$. Introducing a new parameter $\sigma $ instead
of $\tau $, we find 
\begin{equation}
\Bigl(\tau \kappa _{\tau \ast }+v^{2}\kappa _{v\tau \ast }\Bigr)\partial
_{\tau }+v\kappa _{v\ast }\partial _{v}=\sigma \kappa _{\tau \ast }\partial
_{\sigma }+v\kappa _{v\ast }\partial _{v} \ ,
\end{equation}
where
\begin{equation}
\sigma =\tau +a_{\ast }v^{2}\ ,\quad a_{\ast }=\frac{\kappa _{v\tau \ast }}{%
\kappa _{\tau \ast }-2\kappa _{v\ast }}\ .  \label{sigma}
\end{equation}
The $\kappa _{i\ast }$ stand for the Gell-Mann-Low functions taken at the
fixed point $u_{\ast }$.

Using dimensional analysis in conjunction with the flow equations, we readily find
the asymptotic behavior of the connected Green's functions for $l\rightarrow 0$. Neglecting nonuniversal scale factors we get
\begin{eqnarray}
G_{N,\widetilde{N}}(\{{\bf x}\},\sigma ,v,u_{\ast },\mu ) &=&l^{(N+%
\widetilde{N})\eta /2}G_{N,\widetilde{N}}(\{{\bf x}\},\sigma /l^{2-\kappa
_{\tau \ast }},v/l^{1-\kappa _{v\ast }},u_{\ast },\mu l)  \nonumber \\
&=&(\mu l)^{(N+\widetilde{N})(d-2)/2}l^{(N+\widetilde{N})\eta /2}G_{N,%
\widetilde{N}}(\{l\mu {\bf x}\},\sigma /l^{2-\kappa _{\tau \ast
}},v/l^{1-\kappa _{v\ast }},u_{\ast },1)\ ,  \label{G-skal}
\end{eqnarray}
where the Fisher exponent $\eta$ is defined by $\eta =\gamma (u_{\ast })$. We
define the remaining exponents by
\begin{eqnarray}
\nu &=&\frac{1}{1-\kappa _{v\ast }}\ ,\qquad \beta =\nu \frac{d-2+\eta }{2}\
,  \label{nu-beta} \\
\nu _{IP} &=&\frac{1}{2-\kappa _{\tau \ast }}\ ,\qquad \beta _{IP}=\nu _{IP}%
\frac{d-2+\eta }{2}\ ,  \label{iso-nu-beta}
\end{eqnarray}
and the crossover exponent $\phi$ by
\begin{equation}
\phi =\frac{2-\kappa _{\tau \ast }}{1-\kappa _{v\ast }}=\nu /\nu _{IP}=\beta
/\beta _{IP}\ .  \label{phi}
\end{equation}
The appropriate choice of the flow parameter $l$ in the case $\left| \sigma
\right| \ll \left| v\right| ^{\phi }$ is $l\sim \left| v\right| ^{\nu }$. In
this case the Green's functions scale as
\begin{equation}
G_{N,\widetilde{N}}(\{{\bf x}\})=\left| v\right| ^{(N+\widetilde{N})\beta
}F_{N,\widetilde{N}}(\{{\bf x/}\xi \},\sigma /\left| v\right| ^{\phi })\
\label{G-norm}
\end{equation}
with a correlation length $\xi \sim \mu ^{-1}\left| v\right| ^{-\nu }$.
Otherwise, in the case $\left| v\right| \ll \left| \sigma \right| ^{1/\phi }$%
, we choose $l\sim \left| \sigma \right| ^{\nu }$ and arrive at isotropic
percolation scaling
\begin{equation}
G_{N,\widetilde{N}}(\{{\bf x}\})=\left| \sigma \right| ^{(N+\widetilde{N}%
)\beta _{IP}}F_{N,\widetilde{N}}^{(IP)}(\{{\bf x/}\xi _{IP}\},v/\left|
\sigma \right| ^{1/\phi }) \ ,  \label{G-iso}
\end{equation}
with a correlation length $\xi _{IP}\sim \mu ^{-1}\left| \sigma \right|
^{-\nu _{IP}}$.

\section{Results for the Crossover Exponent from the $\varepsilon$-Expansion}

In this section we derive the scaling indices. Because the $\varepsilon$-expansions of the usual percolation exponents are well known, we concentrate on the new crossover exponent $\phi$. For this purpose we need the Gell-Mann-Low functions $\beta _{u}$, $\gamma $, and $\kappa
_{i}$ explicitly. From Eq.~(\ref{betau}) we know that $\beta _{u}$ begins
with the zero-loop term $-\varepsilon u$, the higher order terms are
determined by the Wilson functions. These functions, the logarithmic
derivatives of the $Z$-factors, are given by $\gamma _{...}=\mu \left.
\partial _{\mu }\right| _{0}\ln Z_{...}=\beta _{u}\partial _{u}\ln Z_{...}$.
In minimal renormalization the $Z$-factors have a pure Laurent expansion
with respect to $\varepsilon $: $Z=1+Y^{(1)}/\varepsilon
+Y^{(2)}/\varepsilon ^{2}+\cdots $. It follows then recursively in the loop
expansion that the Wilson functions have also a pure Laurent expansion and,
because they are finite for $\varepsilon \rightarrow 0$, this expansion
reduces to the constant term, i.e., all $\varepsilon $-poles have to be compensated
by the logarithmic derivative. Thus, we get the Wilson functions simply from
the formula $\gamma _{...}=-u\partial _{u}Y_{...}^{(1)}$. Now it is easy to
find these functions from the $Z$-factors Eqs.~(\ref{Z}-\ref{Zu},\ref{Zv},
\ref{Yvt}). The results are
\begin{eqnarray}
\gamma _{u} &=&-4u+\frac{59}{6}u^{2}+O \bigl( u^{3}\bigr) \ ,\quad \gamma
_{\tau }=-u+\frac{47}{24}u^{2}+O \bigl( u^{3}\bigr) \ , \\
\gamma _{v} &=&-\frac{u}{3}+\frac{73}{216}u^{2}+O \bigl(  u^{3}\bigr) \ ,\quad
\gamma _{v\tau }=-\frac{u}{3}+\frac{35}{36}u^{2}+O \bigl(u^{3}\bigr) \ , \\
\gamma &=&-\frac{u}{6}+\frac{37}{216}u^{2}+O\bigl( u^{3}\bigr) \ ,
\end{eqnarray}
from which the Gell-Mann-Low functions follow as
\begin{eqnarray}
\beta _{u} &=&\biggl(-\varepsilon +\frac{7u}{2}-\frac{671}{72}u^{2}+O\bigl(
u^{3}\bigr )\biggr) u\ ,\quad \kappa _{\tau }=\frac{5u}{6}-\frac{193}{108}%
u^{2}+ O \bigl( u^{3}\bigr) \ , \\
\kappa _{v} &=&\frac{u}{6}-\frac{u^{2}}{6}+O\bigl(u^{3}\bigr) \ ,\qquad
\qquad \qquad \quad \kappa _{v\tau }=\frac{u}{3}-\frac{35}{36}u^{2}+O\bigl(
u^{3}\bigr) \ .
\end{eqnarray}
From $\beta_{u\ast }=0$ the stable fixed point value $u_{\ast }=
\frac{2}{7}\varepsilon +\frac{671}{3087}\varepsilon ^{2}+O(\varepsilon ^{3})$ is readily obtained. We finally derive the following $\varepsilon$-expansion of the crossover exponent:
\begin{equation}
\phi =2-\frac{\varepsilon }{7}+\frac{59\varepsilon ^{2}}{2\cdot 21^{3}}+ O
\bigl( \varepsilon ^{3} \bigr) \ .
\end{equation}
Crudely evaluating this expansion for small spatial dimensions, i.e.,\ for $\varepsilon =3$ or $4$, leads inevitably to poor quantitative predictions. Therefore, we improve the $\varepsilon $-expansion via a rational approximation, which takes into account that for $d=1$ the correlation length exponents are, trivially, always equal to one. In addition, we make the hypothesis that the $\varepsilon$-expansion can be extended up to $\varepsilon =5$. These considerations lead then to the interpolation formula
\begin{equation}
\label{ratio}
\phi =1+\biggl(1-\frac{\varepsilon }{5}\biggr)\biggl(1+\frac{2\varepsilon }{
35}+\frac{6767\varepsilon ^{2}}{50\cdot 21^{3}}\biggr)\ .
\end{equation}
From Eq.~(\ref{ratio}) we get values for the crossover and the
order parameter exponent for $d=2$ to $d=6$. These values are summarized in the following table:
\begin{center}
\begin{tabular}{||c||c|c|c|c|c||}
\hline\hline
$\quad d=\quad $ & $2$ & $3$ & $4$ & $5$ & $6$ \\ \hline
$\phi =$ & $\;1.29\;$ & $\;1.52\;$ & $\;1.70\;$ & $\;1.86\;$ & $\;2.00\;$ \\
\hline
$\beta =$ & $\;0.18\;$ & $\;0.62\;$ & $\;1.09\;$ & $\;1.56\;$ & $\;2.00\;$ \\
\hline\hline
\end{tabular}
\end{center}

Of course, the calculated values of $\phi $ depend slightly on the interpolation procedure, i.e., different rational approximands may be used to incorporate $\phi (d=1)=1$. We learn from this numerical sorceries that the displayed numbers may have a failure of roughly $\pm 0.05$.

Now we come back to the question which of the order parameter exponents are
seen in simulations by Inui {\em et al}.~\cite{IKTK99}. From the scaling properties of the Green's functions $G_{N,\widetilde{N}}(\left\{ {\bf x}\right\} )$ we have learned that this
depends on the relative behavior of the relevant parameters $\left| \sigma
\right| $ and $\left| v\right| ^{\phi }$ as functions of the microscopic
probabilities for the conducting elements. Writing the deviations from the
multicritical line as $\delta p=p-p_{c}(r)$, $\delta p_{\pm }=p_{\pm }-r$,
we have the expansions
\begin{eqnarray}
\sigma &=&a_{1}\bigl(\delta p_{+}+\delta p_{-}\bigr)+a_{2}\delta p+a_{3}
\bigl(\delta p_{+}+\delta p_{-}\bigr)^{2}+a_{4}\delta p_{+}\delta
p_{-}+a_{5}\delta p^{2}+...\ , \\
v &=&\bigl(\delta p_{+}-\delta p_{-}\bigr)\Bigl(b_{1}+b_{2}\delta p+b_{3}
\bigl(\delta p_{+}+\delta p_{-}\bigr)\Bigr)... \ ,
\end{eqnarray}
where the coefficients $a_{i},\ b_{i}$ depend on the microscopic
model. In the simulations of the pure two-dimensional random diode
system Inui {\em et al}.~\cite{IKTK99} set $\delta p_{+}=-\delta
p_{-}=\delta r$ and $\delta p=0$. It follows that $\sigma \sim \delta r^{2}$ and $
v\sim \delta r$, and consequently $\left| \sigma \right| /\left| v\right|
^{\phi }\sim \left| \delta r\right| ^{2-\phi }\ll 1$ because $\phi <2$. Thus
the exponent $\beta $ is found as the simulations clearly state. In their
extended model Inui {\em et al}.\ set the variations of the probabilities 
to $\delta p_{+}=\delta p_{-}=\delta r$ and $\delta p=0$. In this case we
have $\sigma \sim \delta r$ and $v\sim \delta r$, and consequently $\left|
v\right| /\left| \sigma \right| ^{1/\phi }\sim \left| \delta r\right|
^{1-1/\phi }\ll 1$ because $1/\phi <1$. Now the isotropic percolation
exponent $\beta _{IP}=5/36$ is measured.

\section{The Crossover from Isotropic to Directed Percolation}

\subsection{Preliminaries}

The crossover theory of Frey, T\"{a}uber, and Schwabl \cite{FTS94},
henceforth called FTS, starts with the quasistatic Hamiltonian ${\cal H}$ stated in
Eq.~(\ref{Hamilt}). FTS use the modification of the longitudinal length scale by
the parameter $c$ which we have discussed in Sec.~II. We have seen that $c$
is a redundant variable near the isotropic fixed point and hence dropped 
it from the onset. FTS renormalize $c$ also. They find that the
anomalous dimension of $c$ is zero at the isotropic point. Moreover, they calculate a nonvanishing anomalous dimension of $c$ at the directed fixed point. The
variable $c$ (or better $1/c^{2}$, which couples to the composed field $
\widetilde{\varphi }\nabla _{\Vert }^{2}\varphi$), being redundant at the isotropic fixed point, changes over to an irrelevant variable at the directed fixed point.
Thus, in order to renormalize it, one has to include all the irrelevant operators of
equal naive dimension and symmetry which mix under renormalization. This was overlooked in the work of FTS, thus the calculated anomalous dimension of $1/c^{2}$ is meaningless.

The technical problem of the crossover from IP to DP is, as FTS stated, that the two
fixed points, which are to connect by the renormalization flow, have different
upper critical dimensions, $6$ and $5$, respectively. Thus, one has from the
outset the problem to renormalize the theory for general dimensions below $6$
and cannot use the $\varepsilon $-expansion, which by-passes the problem
that the perturbation expansion is ill-defined if one uses critical massless
propagators, leading to infrared (IR) divergencies below the upper critical
dimension. However, the renormalization can be accomplished in a massive theory, which avoids the IR divergencies, by using normalization conditions for the vertex
functions. Asymptotic scaling properties follow from the inhomogeneous
Callan-Symanzik equation.

Below the upper critical dimension, the theory is super-renormalizable. Only
the vertex function $\Gamma _{1,1}$ is UV divergent and can be renormalized
by a mass shift $\tau \rightarrow \tau +\delta \tau $, where $\delta \tau $
absorbs the UV divergencies. If the theory is regularized dimensionally,
these UV divergencies manifest themselves as poles at $\varepsilon
=\varepsilon_{l}=2/l$, $l=1,2,...$ \cite{Sym73,BeDa82,BaBe85}, where, at
the corresponding spatial dimensions $d_{l}=6-\varepsilon _{l}$, only
perturbational contributions to $\Gamma _{1,1}$ of loop-order smaller then $
l $ are superficially UV divergent. These poles can be eliminated, as long
as $v$ is finite, by a shift $\delta {\tau }={g}^{4/\varepsilon }{\cal M}
(\varepsilon )$, where ${\cal M}$ is meromorphic in $\varepsilon $. After
the mass shift, the theory is UV-finite. Then, however, in those dimensions
corresponding to the before mentioned poles, non-analytic logarithmic
behavior with respect to the coupling constant arises.

An alternative dimensional regularization formalism was presented some time ago by Schloms and Dohm \cite{SchD89}, henceforth called SD. This formalism circumvents
renormalization conditions and resorts to the more convenient minimal
renormalization but without using the $\varepsilon $-expansion. The key feature of that formalism is to use the inverse correlation length itself as the mass-parameter \cite{BaBe85} and to define the minimal renormalization at a suitable renormalization point, which introduces the usual external mass scale $\mu $. The consequences are relatively simple renormalization factors and a homogeneous RG equation. Hence, the SD formalism fosters higher order calculations and resummation procedures. We
will use the SD formalism in the following. FTS use a massive
renormalization scheme that is a mixture of renormalization conditions and a
mass renormalization at an external mass scale $\mu $. It is unclear wether
this scheme can be used consistently at higher loop-order.

\subsection{Crossover Renormalization}

We revisit the quasistatic Hamiltonian ${\cal H}$, Eqn.~(\ref{Hamilt}), and introduce, in the case $v\neq 0$, new variables by
\begin{equation}
{\bf x}_{\bot }={\bf y\ ,\quad }x_{\Vert }={\bf n\cdot x}=2v\lambda t\
,\quad \varphi =\left| 2v\right| ^{-1/2}s\ ,\quad \widetilde{\varphi }%
=\left| 2v\right| ^{-1/2}\widetilde{s}\ ,\quad g=\left| 2v\right| ^{1/2}\bar{%
g}\ .  \label{DPvar}
\end{equation}
Then ${\cal H}$ appears as
\begin{equation}
\bar{{\cal H}} =\int dtd^{d-1}y\Biggl\{\widetilde{s}\biggl[\partial
_{t}-\frac{1}{4\lambda v^{2}}\partial _{t}^{2}+\lambda \Bigl(\tau -\nabla
^{2}\Bigr)+\frac{\lambda \bar{g}}{2}\Bigl(s-\widetilde{s}\Bigr)\biggr] s
\Biggr\} \ .  \label{DPHam}
\end{equation}
In the limit $v\rightarrow \infty $ we arrive at the dynamic functional
which describes directed percolation (DP) \cite{CaSu80,Ja81}. We conclude
that the so far unrenormalized Green's function have the asymptotic property
\begin{equation}
(2v)^{(N+\widetilde{N})/2}G_{N,\widetilde{N}}\bigl(\{{\bf x}\},\tau
,v,g)\rightarrow \bar{G}_{N,\widetilde{N}}\bigl(\{{\bf x}_{\bot },\lambda
t\},\tau ,\bar{g})\ ,  \label{AsymptGr}
\end{equation}
where the $\bar{G}_{N,\widetilde{N}}$ are the DP Green's functions. The
minimal renormalizations of the DP theory in dimensional regularization are
\begin{eqnarray}
\widetilde{s} &\rightarrow &\mathaccent"7017{\widetilde{s}}=\bar{Z}_{v}^{1/2}%
\widetilde{s}\ ,\quad s\rightarrow \mathaccent"7017{s}=\bar{Z}_{v}^{1/2}s\
,\quad \lambda \rightarrow \mathaccent"7017{\lambda }=\bar{Z}_{v}^{-1}\bar{Z}%
\lambda \ ,  \label{DPRena} \\
\tau &\rightarrow &\mathaccent"7017{\tau }=\bar{Z}^{-1}\bar{Z}_{\tau }\tau
\mu ^{2}+\mathaccent"7017{\bar{\tau}}_{c}\ ,\quad \bar{g}^{2}\rightarrow %
\mathaccent"7017{\bar{g}}^{2}=\bar{G}_{\bar{\varepsilon}}^{\,-1}\bar{Z}%
_{v}^{-1}\bar{Z}^{-2}\bar{Z}_{\bar{u}}\bar{u}\mu ^{\bar{\varepsilon}}\ ,
\label{DPRenb}
\end{eqnarray}
where $\bar{\varepsilon}=\varepsilon -1=5-d=4-\bar{d}$, $\bar{G}_{\bar{%
\varepsilon}}=\Gamma (1+\bar{\varepsilon}/2)/(4\pi )^{\bar{d}/2}$, and known
\cite{Ja81} as
\begin{eqnarray}
\bar{Z} &=&1+\frac{\bar{u}}{8\bar{\varepsilon}}+\frac{\bar{u}^{2}}{%
128\varepsilon }\biggl(\frac{13}{\varepsilon }-\frac{31}{4}+\frac{35}{2}\ln
\frac{4}{3}\biggr)+O(\bar{u}^{3})\ ,  \label{ZDP} \\
\bar{Z}_{v} &=&1+\frac{\bar{u}}{4\bar{\varepsilon}}+\frac{\bar{u}^{2}}{%
32\varepsilon }\biggl(\frac{7}{\varepsilon }-3+\frac{9}{2}\ln \frac{4}{3}%
\biggr)+O(\bar{u}^{3})\;,  \label{ZvDP} \\
\bar{Z}_{\tau } &=&1+\frac{\bar{u}}{2\bar{\varepsilon}}+\frac{\bar{u}^{2}}{%
2\varepsilon }\biggl(\frac{1}{\varepsilon }-\frac{5}{16}\biggr)+O(\bar{u}%
^{3})\ ,  \label{ZtDP} \\
\bar{Z}_{u} &=&1+\frac{2\bar{u}}{\bar{\varepsilon}}+\frac{7\bar{u}^{2}}{%
2\varepsilon }\biggl(\frac{1}{\varepsilon }-\frac{1}{4}\biggr)+O(\bar{u}%
^{3})\ .  \label{ZuDP}
\end{eqnarray}
Note that the DP-Hamiltonian ${\bar{{\cal H}}}$, Eq.~(\ref{DPHam}), is non-renormalizable for $d>5$. Thus, to study this limit we are restricted to spatial dimensions $d\leq 5$. 

We expect that the theory can be rendered finite for all values of $v$,
including infinity, by interpolating the renormalizations of ${\cal H}$. Then it follows from the definitions~(\ref{DPvar}) that the following asymptotic properties hold for $v\rightarrow \infty$:
\begin{equation}
Z_{...}\bigl(u,v,\varepsilon \bigr)\rightarrow \bar{Z}_{...}\bigl(\bar{u},
\bar{\varepsilon}\bigr)\ ,  \label{Zasympt}
\end{equation}
with $\bar{u}=u/2v$. Of course, as long as $v$ is finite the soft
renormalizations of ${\cal H}$ can be used, but for $v\rightarrow \infty $,
logarithmic divergencies $\sim \ln v$ arise for $\bar{\varepsilon}
=\varepsilon -1=0$. These must be included in the renormalization factors to
make the theory finite for all $v$, especially in the DP limit. Note that in
this limit $v$ plays the role of a cutoff. In the following we demonstrate
that the renormalization procedure suggested by SD, i.e., an extended minimal renormalization scheme without using the $\varepsilon$-expansion, is feasible and that it allows to extract the crossover behavior of the Green's functions.

We consider the dimensional regularized bare vertex functions $\mathaccent
"7017{\Gamma }_{\widetilde{N},N}\bigl(\{{\bf q}\};\mathaccent"7017{\tau },
\mathaccent"7017{v},\mathaccent"7017{g};d\bigr)$ as functions of the bare
parameters $\mathaccent"7017{\tau },\mathaccent"7017{v},\mathaccent"7017{g}$
, and the momenta $\{{\bf q}\}$ at spatial dimension $d=6-\varepsilon $, $
\varepsilon >0$. The critical point $\mathaccent"7017{\tau }_{c}$ is
determined by
\begin{equation}
\mathaccent"7017\Gamma _{1,1}\bigl(\{{\bf 0}\};\mathaccent"7017\tau _{c},
\mathaccent"7017v,\mathaccent"7017g;d\bigr)=0\ ,
\end{equation}
which provides an implicit definition of
\begin{equation}
\mathaccent"7017\tau _{c}=\mathaccent"7017\tau _{c}\bigl(\mathaccent"7017v,
\mathaccent"7017g;d\bigr)=\mathaccent"7017g^{4/\varepsilon }S\bigl(
\varepsilon ,\mathaccent"7017v/\mathaccent"7017g^{2/\varepsilon }\bigr)\ .
\end{equation}
The generalized Symanzik function \cite{Sym73} has pole singularities at $
\varepsilon =2/l$, $l=1,2,...$, as long as $x=\mathaccent"7017v/\mathaccent
"7017g^{2/\varepsilon }$ is finite. An expansion yields $S(\varepsilon
,x)=S_{0}(\varepsilon )+Y(\varepsilon )x^{2}+O(x^{4})$ with a finite $
Y(\varepsilon )$ as long as $\varepsilon >0$. Taking  the limit behavior of the vertex functions for $x\gg 1$ into account, we find for $\varepsilon >1$, the
asymptotic behavior $S(\varepsilon ,x)\rightarrow x^{2-\varepsilon }\bar{P}
(x^{-\varepsilon };\varepsilon )+x^{2/(1-\varepsilon )}\bar{S}(\varepsilon )$, where $\bar{P}$ denotes an analytic function of $x^{-\varepsilon }=
\mathaccent"7017g^{2}/\mathaccent"7017v^{\varepsilon }$. For $\varepsilon >1$
, the pole-singularities of $\bar{P}(y;\varepsilon )$ and $\bar{S}
(\varepsilon )$ are found at $\varepsilon -1=\bar{\varepsilon}=2/l$. These pole-singularities combine to yield logarithmic divergencies in $x$ instead the poles in $\bar{\varepsilon}$.

Since the theory is super-renormalizable for $\varepsilon >0$, we know that
all the functions 
\begin{equation}
\mathaccent"7017\Gamma _{\widetilde{N},N}^{\prime }\bigl(
\{{\bf q}\};\mathaccent"7017\tau ^{\prime },\mathaccent"7017v,\mathaccent
"7017g;d\bigr)=\mathaccent"7017\Gamma _{\widetilde{N},N}\bigl(\{{\bf q}\};
\mathaccent"7017\tau ^{\prime }+\mathaccent"7017\tau _{c}\bigl(\mathaccent
"7017v,\mathaccent"7017g;d\bigr),\mathaccent"7017v,\mathaccent"7017g;d\bigr)
\end{equation}
are finite below six dimensions. However, because of the non-analyticity of $
\mathaccent"7017\tau _{c}$ with respect to $\mathaccent"7017{g}$, they are
no more expandable in the coupling constant $\mathaccent"7017g$. This can
be fixed by introducing the inverse transversal correlation length as the
new mass. It is defined according to
\begin{equation}
\xi _{\bot }^{-2}=m^{2}=\frac{\mathaccent"7017\Gamma _{1,1}^{\prime }\bigl(\{
{\bf 0}\};\mathaccent"7017{\tau}^{\prime},\mathaccent"7017v,\mathaccent
"7017g;d\bigr)}{\left. \partial _{q_{\bot }^{2}}\mathaccent"7017\Gamma
_{1,1}^{\prime }\bigl(\{{\bf q}\};\mathaccent"7017\tau ^{\prime },\mathaccent
"7017v,\mathaccent"7017g;d\bigr)\right| _{{\bf q}=0}}\ .  \label{mass}
\end{equation}
The function $m\bigl(\mathaccent"7017\tau ^{\prime },\mathaccent"7017v,
\mathaccent"7017g;d\bigr)$ has poles only at $d=6$. It can be inverted to
define $\mathaccent"7017\tau ^{\prime }$ as a function of $m$,
\begin{equation}
\mathaccent"7017\tau ^{\prime }=\mathaccent"7017r\bigl(m,\mathaccent"7017v,
\mathaccent"7017g;d\bigr)\ ,  \label{rmass}
\end{equation}
with $\mathaccent"7017r\bigl(0,\mathaccent"7017v,\mathaccent"7017g;d\bigr)=0$. The function $\mathaccent"7017r$ can be substituted into $\mathaccent
"7017\Gamma _{\widetilde{N},N}^{\prime }$ yielding the bare vertex functions $\mathaccent"7017\Gamma _{
\widetilde{N},N}^{\prime \prime }$ in terms of the mass $m$:
\begin{equation}
\mathaccent"7017\Gamma _{\widetilde{N},N}^{\prime }\bigl(\{{\bf q}\};
\mathaccent"7017r\bigl(m,\mathaccent"7017v,\mathaccent"7017g;d\bigr),
\mathaccent"7017v,\mathaccent"7017g;d\bigr)=\mathaccent"7017\Gamma _{
\widetilde{N},N}^{\prime \prime }\bigl(\{{\bf q}\};m,\mathaccent"7017v,
\mathaccent"7017g;d\bigr)\ .
\end{equation}
In the following we abbreviate $\mathaccent"7017\Gamma _{\widetilde{N},N}^{\prime \prime }$ by $\mathaccent"7017\Gamma _{\widetilde{N},N}$ for notational simplicity. The
$\mathaccent"7017\Gamma _{\widetilde{N},N}$ are now free of dimensional singularities below $d=6$, and have an expansion in integer powers of $\mathaccent"7017g$. In the limit $v\rightarrow \infty $, for $d<5$, we deduce from Eq.\ (\ref{AsymptGr}) the
asymptotic property
\begin{equation}
\mathaccent"7017\Gamma _{\widetilde{N},N}\bigl(\{{\bf q}\},m,\mathaccent
"7017v,\mathaccent"7017g;d)\rightarrow (2\mathaccent"7017v)^{(N+\widetilde{N}
-2)/2}\mathaccent"7017{\bar{\Gamma}}_{\widetilde{N},N}\bigl(\{{\bf q}_{\bot
},\mathaccent"7017\lambda ^{-1}\omega \},m,\mathaccent"7017{\bar{g}};d)\ ,
\label{AsymptVert}
\end{equation}
with $\omega =2\mathaccent"7017\lambda \mathaccent"7017vq_{\Vert }$ and the
unrenormalized DP vertex functions $\mathaccent"7017{\bar{\Gamma}}_{
\widetilde{N},N}$.

Henceforth we use for simplicity the notation $q$ symbolically for all
the momenta $\{{\bf q}\}$. It is convenient to write
\begin{equation}
\mathaccent"7017\Gamma _{\widetilde{N},N}\bigl(q,m,\mathaccent"7017v,
\mathaccent"7017g;d\bigr)=m^{\delta _{_{\widetilde{N},N}}}\mathaccent"7017{F}
_{\widetilde{N},N}\bigl(q/m,\mathaccent"7017v/m,\mathaccent
"7017g/m^{\varepsilon /2};d\bigr)
\end{equation}
with dimensionless functions $\mathaccent"7017{F}_{\widetilde{N},N}$ and $
\delta _{_{\widetilde{N},N}}=d-(d-2)\bigl(\widetilde{N}+N\bigr)/2$. Expanding $\mathaccent"7017{F}_{\widetilde{N},N}$ in a power series gives
\begin{equation}
\mathaccent"7017{F}_{\widetilde{N},N}\bigl(x,y,z;d\bigr)=z^{\sigma
}\sum_{l,n=0}^{\infty }f_{\widetilde{N},N}^{(l,n)}\bigl(y;d\bigr)\bigl( x
\bigr)^{n}\Bigl(z^{2}/\varepsilon \Bigr)^{l}\ ,  \label{SkalVert}
\end{equation}
where $\sigma =0$ if $\bigl(\widetilde{N}+N\bigr)/2$ is an integer and $
\sigma =1$ if not. The exponent $l$ denotes the loop order. The functions $
f_{\widetilde{N},N}^{(l,n)}\bigl(y;d\bigr)$ are finite for $d\leq 6$ and
finite $y$. Moreover, they are analytic in $y$ and can be expanded to
\begin{equation}
f_{\widetilde{N},N}^{(l,n)}\bigl(y;d\bigr)=\sum_{k=0}^{\infty }f_{\widetilde{
N},N}^{(k,l,n)}\bigl(d\bigr)y^{k}
\end{equation}

The contributions of order $z^{l}$ to the scaled vertex functions $
\mathaccent"7017{F}_{\widetilde{N},N}$ (\ref{SkalVert}) exhibit infrared
(IR) divergencies for $m\rightarrow 0$. SD showed at the instance of the $\Phi ^{4}$-theory, that this problem can be treated by minimal renormalizations
\begin{eqnarray}
\mathaccent"7017{\widetilde{\varphi }} &=&Z^{1/2}\widetilde{\varphi }\
,\quad \mathaccent"7017{\varphi }=Z^{1/2}\varphi \ ,  \label{Ren2a} \\
\mathaccent"7017{g}^{2} &=&A\bigl(d\bigr)G_{\varepsilon
}^{\,-1}Z^{-3}Z_{u}u\mu ^{\varepsilon }\ ,\quad v\rightarrow \mathaccent"7017
{v}=Z^{-1}Z_{v}v\mu \ ,  \label{Ren2b}
\end{eqnarray}
where the soft renormalization factors $Z_{...}\bigl(u;d\bigr)$ absorb just
the poles at $d=6$, followed by the application of the renormalization
group. The UV $\varepsilon $-poles and the IR divergencies are then summed
up yielding the correct critical scaling behavior reminiscent of the intimate
relation between both. Since the poles of the coefficients $f_{\widetilde{N}
,N}^{(k,l,n)}\bigl(6-\varepsilon \bigr)/\varepsilon ^{l}$ do not depend on
any of the parameters, they are identical to the usual soft minimal
renormalizations and can be calculated by setting $m=\mu $ or in the
massless theory with $\varepsilon \rightarrow 0$. Following SD, we have
introduced in Eq.~(\ref{Ren2b}) a further amplitude $A(d)$ with $A(0)=1$. This dimension
dependent function can be conveniently defined and is extremely useful for
the practical calculation of scaling functions. For simplicity we set $
A(d)=1$ in the remaining part of this article. Note that from its definition
(\ref{mass}), the mass $m$ does not need a multiplicative renormalization and that $m=
\mathaccent"7017m$.

Now we consider the DP vertex functions
\begin{equation}
\mathaccent"7017{\bar{\Gamma}}_{\widetilde{N},N}\bigl(q_{\bot },\mathaccent
"7017\lambda ^{-1}\omega ,m,\mathaccent"7017{\bar{g}};d\bigr)=m^{\bar{\delta}
_{_{\widetilde{N},N}}}\mathaccent"7017{\bar{F}}_{\widetilde{N},N}\bigl(
q_{\bot }/m,\omega /\mathaccent"7017\lambda m^{2},\mathaccent"7017{\bar{g}}
/m^{\bar{\varepsilon}/2};d\bigr)\ ,  \label{DPVert}
\end{equation}
with $\delta _{_{\widetilde{N},N}}=(d+1)-(d-1)\bigl(\widetilde{N}+N\bigr)/2$
. The power series expansion of the $\mathaccent"7017{\bar{F}}_{\widetilde{N}
,N}$ reads for $d<5$
\begin{equation}
\mathaccent"7017{\bar{F}}_{\widetilde{N},N}\bigl(x_{\bot },\bar{x}_{\Vert },
\bar{z};d\bigr)=\bar{z}^{\sigma }\sum_{l,n_{\bot },n_{\Vert }=0}^{\infty }
\bar{f}_{\widetilde{N},N}^{(l,n_{\bot },n_{\Vert })}\bigl(\bar{d}\bigr)\bigl(
x_{\bot }\bigr)^{n_{\bot }}\bigl(x_{\Vert }\bigr)^{n_{\Vert }}\Bigl(\bar{z}
^{2}/\bar{\varepsilon}\Bigr)^{l}\ ,  \label{SkalDPVert}
\end{equation}
with $\bar{\varepsilon}=5-d=4-\bar{d}$. By the same argumentation as above,
one can show that the $\bar{\varepsilon}$-poles are cancelled and the IR
divergencies are summed up by the renormalization group equation based on
the minimal renormalizations $\bar{Z}$, $\bar{Z}_{v}$, and $\bar{Z}_{u}$,
Eqs.~(\ref{ZDP},\ref{ZvDP},\ref{ZuDP}), where the renormalization factors $
\bar{Z}_{...}\bigl(\bar{u},d\bigr)$ now absorb just the $\bar{\varepsilon}$
-poles at $d=5$. As above, they are identical to the usual soft minimal
renormalizations. They can be calculated at the renormalization point $m=\mu $
, $\{{\bf q}=0\}$ or in the massless theory with $\bar{\varepsilon}
\rightarrow 0$, because the functions $\bar{f}_{\widetilde{N},N}^{(l,n_{\bot
},n_{\Vert })}\bigl(\bar{d}\bigr)$ do not depend on any of the parameters.

Combining the asymptotic properties of the vertex functions for $
y\rightarrow \infty $, Eq.\ (\ref{AsymptVert}), with the various expansions,
we find for $d<5$
\begin{equation}
f_{\widetilde{N},N}^{(l,n)}\bigl(y;d\bigr)\rightarrow (2y)^{(N+\widetilde{N}
-2-\sigma )/2+n_{\Vert }-l}\Bigl(\varepsilon /\bar{\varepsilon}\Bigr)^{l}
\bar{f}_{\widetilde{N},N}^{(l,n_{\bot },n_{\Vert })}\bigl(\bar{d}\bigr) \ .
\end{equation}
In case of taking the limit $d\rightarrow 5$ before $y\rightarrow \infty $,
the $\bar{\varepsilon}^{-l}$-poles are replaced by a polynomial $
P^{(l,n_{\bot },n_{\Vert })}(\ln y)$ of order $l$. Therefore, it should be 
possible to find further finite renormalizations $\zeta _{...}\bigl(u,v;d
\bigr)$ which interpolate between the two minimal renormalizations at the
renormalization point $m=\mu$,
\begin{equation}
Z_{...}\bigl(u;d\bigr)\rightarrow \zeta _{...}\bigl(u,v;d\bigr)Z_{...}\bigl(
u;d\bigr)=Z_{...}\bigl(u,v;d\bigr)\ ,  \label{CrossRen}
\end{equation}
where the $Z_{...}\bigl(u,v;d\bigr)$ tend to the corresponding $\bar{Z}_{...}
\bigl(\bar{u};\bar{d}\bigr)$ with $\bar{u}=u/2v$ in the limit $v\rightarrow
\infty $.

Following SD, the remaining problem is to determine the renormalization of the
relation between the mass $m$ and the ``temperature'' $\mathaccent"7017{\tau
}=\mathaccent"7017{\tau }\bigl(m,\mathaccent"7017{v},\mathaccent"7017{g};d
\bigr)=m^{2}\mathaccent"7017{T}\bigl(\mathaccent"7017{v}/m,\mathaccent"7017{g
}/m^{\varepsilon };d\bigr)$. The dimensionless function $\mathaccent"7017{T}$
has an expansion
\begin{equation}
\mathaccent"7017{T}\bigl(y,z;d\bigr)=1+\sum_{l=1}^{\infty }t_{l}\bigl(y;d
\bigr)\Bigl(z^{2}/\varepsilon \Bigr)^{l}\ ,
\end{equation}
but the $t_{l}(y;d)$ show UV singularities below $d=6$. Otherwise, the use
of the variable $\mathaccent"7017{\tau }^{\prime }=\mathaccent"7017{\tau }
\mathaccent"7017{\tau }_{c}=\mathaccent"7017{r}\bigl(m,\mathaccent"7017{v},
\mathaccent"7017{g};d\bigr)=$ $m^{2}\mathaccent"7017{T}\bigl(\mathaccent"7017
{v}/m,\mathaccent"7017{g}/m^{\varepsilon };d\bigr)-\mathaccent"7017{g}
^{4/\varepsilon }S\bigl(\varepsilon ,v/\mathaccent"7017{g}^{2/\varepsilon }
\bigr)$ instead of $\mathaccent"7017{\tau }$ eliminates all UV singularities
below the upper critical dimension. Then, however, the function $\mathaccent"7017{r
}$ is not more expandable in $\mathaccent"7017{g}$. Following SD, we shall
therefore consider the derivative $\left. \partial \mathaccent"7017{r}
/\partial m^{2}\right| _{\mathaccent"7017{v},\mathaccent"7017{g}}$, which is
not only expandable but also free of singularities. We define the function
\begin{equation}
\mathaccent"7017{P}\bigl(\mathaccent"7017{v}/m,\mathaccent"7017{g}
/m^{\varepsilon };d\bigr)=\left. \frac{\partial \mathaccent"7017{r}}{
\partial m^{2}}\right| _{\mathaccent"7017{v},\mathaccent"7017{g}}=\left.
\frac{\partial \bigl(m^{2}\mathaccent"7017{T}\bigr)}{\partial m^{2}}\right|
_{\mathaccent"7017{v},\mathaccent"7017{g}}=\left. \frac{\partial \mathaccent
"7017{\tau }}{\partial m^{2}}\right| _{\mathaccent"7017{v},\mathaccent"7017{g
}}  \label{PDef}
\end{equation}
which has the expansion
\begin{equation}
\mathaccent"7017{P}\bigl(y,z;d\bigr)=1+\sum_{l=1}^{\infty }p_{l}\bigl(y;d
\bigr)\Bigl(z^{2}/\varepsilon \Bigr)^{l}\ .
\end{equation}
The functions $p_{l}\bigl(y;d\bigr)$ are finite for $d\leq 6$, and $
\mathaccent"7017{P}$ is minimally renormalized by
\begin{equation}
\mathaccent"7017{P}=Z^{-1}Z_{\tau }P\ .  \label{PRen}
\end{equation}
The same argumentation as above shows that it is possible to find a $Z_{\tau
}\bigl(u,v;d\bigr),$ which interpolates between the respective minimal
renormalizations of isotropic and directed percolation. Note that $\left.
\partial /\partial m^{2}\right| _{\mathaccent"7017{v},\mathaccent"7017{g}
}=\left. \partial /\partial m^{2}\right| _{{v},{g,\mu }}$. Thus, the
derivative with respect to the mass $m$ commutes with the multiplication with $Z$-factors.

\subsection{One-Loop Crossover Calculation}

In this subsection we derive a minimal crossover renormalization, the
corresponding renormalization group equation, and the flow equations, which
show the crossover from isotropic to directed percolation. From the
self-energy to order $g^{2}$, we get the unrenormalized (for notational simplicity we drop the overcirc) vertex function $\Gamma _{1,1}$ expanded to
second order in the momenta:
\begin{eqnarray}
\Gamma _{1,1} &=&\tau \biggl(1-\frac{2G_{\varepsilon }g^{2}\tau
^{-\varepsilon /2}}{(2-\varepsilon )\varepsilon }K_{\varepsilon -4}^{(0)}
\bigl(v/\sqrt{\tau }\bigr)\biggr)+2i{\bf v\cdot q}\biggl(1-\frac{
G_{\varepsilon }g^{2}\tau ^{-\varepsilon /2}}{2\varepsilon }K_{\varepsilon
-2}^{(1)}\bigl(v/\sqrt{\tau }\bigr)\biggr)  \nonumber \\
&&+q^{2}\biggl(1-\frac{G_{\varepsilon }g^{2}\tau ^{-\varepsilon /2}}{
4\varepsilon }K_{\varepsilon -2}^{(1)}\bigl(v/\sqrt{\tau }\bigr)\biggr)-
\bigl({\bf v\cdot q}\bigr)^{2}\frac{G_{\varepsilon }g^{2}\tau ^{-\varepsilon
/2}}{4\tau }K_{\varepsilon }^{(2)}\bigl(v/\sqrt{\tau }\bigr)\ .
\label{Vert11Cr}
\end{eqnarray}
The inverse transversal correlation length is given by
\begin{equation}
\label{keineIdee}
m^{2}=\frac{\tau \Bigl(1-\frac{2G_{\varepsilon }g^{2}\tau ^{-\varepsilon /2}
}{(2-\varepsilon )\varepsilon }K_{\varepsilon -4}^{(0)}\bigl(v/\sqrt{\tau }
\bigr)\Bigr)}{\Bigl(1-\frac{G_{\varepsilon }g^{2}\tau ^{-\varepsilon /2}}{
4\varepsilon }K_{\varepsilon -2}^{(1)}\bigl(v/\sqrt{\tau }\bigr)\Bigr)}\ .
\end{equation}
From Eq.~(\ref{keineIdee}) we find easily the unrenormalized perturbational temperature as
\begin{equation}
\tau =m^{2}\biggl(1+\frac{G_{\varepsilon }g^{2}m^{-\varepsilon }}{
4\varepsilon }\Bigl(\frac{8}{(2-\varepsilon )}K_{\varepsilon
-4}^{(0)}(v/m)-K_{\varepsilon -2}^{(1)}(v/m)\Bigr)\biggr)\ .  \label{t-mass}
\end{equation}

Taking the derivative with respect to $m^{2}$ while holding $g$ and $v$ constant and renormalization by multiplication with the factor $ZZ_{\tau }^{-1}$ yields via the renormalization scheme the function $P$ as
\begin{equation}
P=ZZ_{\tau }^{-1}+u(\mu /m)^{\varepsilon }\biggl(\frac{1}{4\varepsilon }
\Bigl(4K_{\varepsilon -2}^{(0)}(v\mu /m)-K_{\varepsilon -2}^{(1)}(v\mu
/m)\Bigr)-\frac{1}{8}K_{\varepsilon }^{(1)}(v\mu /m)\biggr)\ .  \label{PRen1}
\end{equation}
With the formulas
\begin{eqnarray}
p^{2}K_{\alpha }^{(1)}(p) &=&\bigl(1+p^{2}\bigr)K_{\alpha }(p)-K_{\alpha
-2}(p)\ ,  \label{Kpred1} \\
(1+\alpha )K_{\alpha }(p) &=&(2+\alpha )K_{\alpha +2}(p)-\bigl(1+p^{2}\bigr)
^{-1-\alpha /2} \ , \label{Kpred2}
\end{eqnarray}
where we have defined $K_{\alpha }:=K_{\alpha }^{(0)}$, we get 
\begin{eqnarray}
P &=&ZZ_{\tau }^{-1}+\frac{u}{8}\Biggl(\frac{20}{3\varepsilon }\bigl(1+v^{2}
\bigr)^{-\varepsilon /2}+\frac{2}{\varepsilon -1}\bigl(3-v^{-2}\bigr)\biggl[
K_{\varepsilon }(v)-\bigl(1+v^{2}\bigr)^{-\varepsilon /2}\biggr]  \nonumber
\\
&&+\frac{1}{\varepsilon -3}\biggl(v^{-2}\biggl[K_{\varepsilon }(v)-\bigl(
1+v^{2}\bigr)^{-\varepsilon /2}\biggr]-\frac{2}{3}\bigl(1+v^{2}\bigr)
^{-\varepsilon /2}\biggr)+K_{\varepsilon }(v)\Biggr)  \label{PRen2}
\end{eqnarray}
at the renormalization point $m=\mu $. Note that the poles for $\varepsilon =1$ and $\varepsilon =3$ are only fictitious because $K_{1}(p)=(1+p^{2})^{-1/2}$ and $
K_{3}(p)=(1+p^{2})^{-3/2}(1+2p^{2}/3)$. Useful properties of the function $
K_{\varepsilon }(p)$ are
\begin{eqnarray}
K_{\varepsilon }(p) &=&1-\frac{2+\varepsilon }{6}p^{2}+O\Bigl(p^{4}\Bigr)\
\label{Kpklein} \\
&=&\frac{\sqrt{\pi }\Gamma ((1+\varepsilon )/2)}{2\Gamma (1+\varepsilon /2)}
p^{-1}-\frac{1}{1+\varepsilon }p^{-2-\varepsilon }+O\Bigl(p^{-4-\varepsilon
}\Bigr)\ ,  \label{Kpgross}
\end{eqnarray}
from which we find that the fictitious pole at $\varepsilon =1$ in Eq.\ (\ref
{PRen2}) is replaced by a divergence $\sim \ln v$, and that $w=uK_{\varepsilon
}(v)\rightarrow \bar{u}/2$ for $v\rightarrow \infty $. Thus, a minimal
crossover renormalization is given by
\begin{equation}
ZZ_{\tau }^{-1}=1-\frac{5u}{6\varepsilon }\bigl(1+v^{2}\bigr)^{-\varepsilon
/2}-\frac{3u}{4(\varepsilon -1)}\biggl[K_{\varepsilon }(v)-\bigl(1+v^{2}
\bigr)^{-\varepsilon /2}\biggr]+O\bigl(u^{2}\bigr)\ .  \label{ZZt}
\end{equation}
Defining the crossover function
\begin{equation}
C_{\varepsilon }(v)=1-\Bigl(\bigl(1+v^{2}\bigr)^{\varepsilon
/2}K_{\varepsilon }(v)\Bigr)^{-1}\ ,  \label{Cfunkt}
\end{equation}
we render the renormalized temperature derivative
\begin{equation}
P=1+\frac{w}{8}\biggl(1+\frac{1}{3-\varepsilon }\Bigl(2\frac{
1-C_{\varepsilon }(v)}{3}+\frac{5-\varepsilon }{1-\varepsilon }
v^{-2}C_{\varepsilon }(v)\Bigr)\biggr)  \label{Pfunkt}
\end{equation}
finite for all $d\leq 5$ and all $v$.

Now we consider the vertex function $\Gamma _{1,1}$, Eq.\ (\ref{Vert11Cr}),
for ${\bf q}\neq 0$. Inserting $\tau $, Eq.\ (\ref{t-mass}), we get after
renormalization
\begin{eqnarray}
\Gamma _{1,1} &=&\bigl(m^{2}+q^{2}\bigr)\biggl(Z-\frac{u(\mu
/m)^{\varepsilon }}{4\varepsilon }K_{\varepsilon -2}^{(1)}\bigl(v\mu /m\bigr)
\biggr)  \nonumber \\
&&+2i\mu {\bf v\cdot q}\biggl(Z_{v}-\frac{u(\mu /m)^{\varepsilon }}{
2\varepsilon }K_{\varepsilon -2}^{(1)}\bigl(v\mu /m\bigr)\biggr)-\bigl({\bf 
v\cdot q}\bigr)^{2}\frac{u(\mu /m)^{2+\varepsilon }}{4}K_{\varepsilon }^{(2)}
\bigl(v\mu /m\bigr)\ .  \label{Vert11CrRen}
\end{eqnarray}
The last term is finite for all $d\leq 5$ and all $v$, even in the limit $
v\rightarrow \infty $, if one holds $2{\bf v\cdot q=}\omega /\lambda $ and $
w=uK_{\varepsilon }(v)$ constant. The reduction of $K_{\varepsilon
-2}^{(1)}(v)$ with Eqs.\ (\ref{Kpred1},\ref{Kpred2}) leads to
\begin{eqnarray}
K_{\varepsilon -2}^{(1)}(v) &=&\frac{2}{3\varepsilon }\bigl(1+v^{2}\bigr)
^{-\varepsilon /2}+\frac{2+v^{-2}}{2(\varepsilon -1)}\biggl[K_{\varepsilon
}(v)-\bigl(1+v^{2}\bigr)^{-\varepsilon /2}\biggr]  \nonumber \\
&&-\frac{1}{\varepsilon -3}\biggl(\frac{1}{2}v^{-2}\biggl[K_{\varepsilon
}(v)-\bigl(1+v^{2}\bigr)^{-\varepsilon /2}\biggr]-\frac{1}{3}\bigl(1+v^{2}
\bigr)^{-\varepsilon /2}\biggr)\ .  \label{K1p}
\end{eqnarray}
We see that the divergencies which have to be absorbed by renormalization
arise from the first two terms of the right hand side of Eq.~(\ref{K1p}). Thus, we define
\begin{eqnarray}
Z &=&1+\frac{u}{6\varepsilon }\bigl(1+v^{2}\bigr)^{-\varepsilon /2}+\frac{u}{
4(\varepsilon -1)}\biggl[K_{\varepsilon }(v)-\bigl(1+v^{2}\bigr)
^{-\varepsilon /2}\biggr]+O\bigl(u^{2}\bigr)\ ,  \label{Zcross} \\
Z_{v} &=&1+\frac{u}{3\varepsilon }\bigl(1+v^{2}\bigr)^{-\varepsilon /2}+
\frac{u}{2(\varepsilon -1)}\biggl[K_{\varepsilon }(v)-\bigl(1+v^{2}\bigr)
^{-\varepsilon /2}\biggr]+O\bigl(u^{2}\bigr)\ .  \label{Zvcross}
\end{eqnarray}

Now we turn to the calculation of the one-loop contribution to the vertex
function $\Gamma _{1,2}$. To renormalize $\Gamma _{1,2}$ we need it for zero
external momenta. We find the contribution
\begin{equation}
V^{(1)}=-2g^{3}\int_{{\bf k}}G({\bf k})^{2}G(-{\bf k})=2g\frac{\partial }{
\partial \tau }\Sigma ^{(1)}({\bf 0})=\frac{2G_{\varepsilon }g^{3}}{
\varepsilon }\tau ^{-\varepsilon /2}K_{\varepsilon -2}\bigl(v/\sqrt{\tau }
\bigr)\ ,
\end{equation}
which leads after renormalization to
\begin{equation}
\bigl(\Gamma _{1,2}\bigr)^{2}=G_{\varepsilon }^{-1}u\mu ^{\varepsilon }
\biggl(Z_{u}-\frac{4u(\mu /m)^{\varepsilon }}{\varepsilon }K_{\varepsilon -2}
\bigl(v\mu /m\bigr)\biggr)\ .
\end{equation}
Using once more the reduction formula (\ref{Kpred2}), we finally find 
\begin{equation}
Z_{u}=1+\frac{4u}{\varepsilon }\bigl(1+v^{2}\bigr)^{-\varepsilon /2}+\frac{4u
}{(\varepsilon -1)}\biggl[K_{\varepsilon }(v)-\bigl(1+v^{2}\bigr)
^{-\varepsilon /2}\biggr]+O\bigl(u^{2}\bigr)\ .  \label{Zucross}
\end{equation}

Now, all the $Z$-factors (\ref{ZZt},\ref{Zcross},\ref{Zvcross},\ref{Zucross}
) are of the form
\begin{equation}
Z_{i}=1+u\biggl(\frac{a_{i}}{\varepsilon }\bigl(1+v^{2}\bigr)^{-\varepsilon
/2}+\frac{b_{i}}{(\varepsilon -1)}\biggl[K_{\varepsilon }(v)-\bigl(1+v^{2}
\bigr)^{-\varepsilon /2}\biggr]\biggr)+O\bigl(u^{2}\bigr)\ ,
\end{equation}
where the coefficients are given by
\begin{eqnarray}
a &=&1/6\ ,\quad b=1/4\ ,\qquad a_{\tau }=1\ ,\quad b_{\tau }=1\ ,
\label{Koeffa} \\
a_{v} &=&1/3\ ,\quad b_{v}=1/2\ ,\qquad a_{u}=4\ ,\quad b_{u}=4\ .
\label{Koeffb}
\end{eqnarray}
With help of the derivative of the $K$-function,
\begin{equation}
v\frac{\partial K_{\varepsilon }(v)}{\partial v}=\bigl(1+v^{2}\bigr)
^{-1-\varepsilon /2}-K_{\varepsilon }(v)=-K_{\varepsilon }(v)C_{\varepsilon
}(v) \ ,
\end{equation}
and the Gell-Mann-Low functions 
\begin{eqnarray}
\beta _{u} &=& \left. \partial
u/\partial \ln \mu \right| _{0}=(-\varepsilon +3\gamma -\gamma
_{u})u=-\varepsilon u+O(u^{2}) \ ,
\\
 \beta _{v} &=& \left. \partial u/\partial \ln
\mu \right| _{0}=(-1+\gamma -\gamma _{v})v=(-1+O(u))v
\end{eqnarray}
we obtain the Wilson-functions $\gamma _{i}=\left. \partial \ln Z_{i}/\partial \ln \mu
\right| _{0}=\left. (\beta _{u}\partial _{u}+\beta _{v}\partial _{v})\ln
Z_{i}\right| _{0}$ easily as
\begin{eqnarray}
\gamma _{i} &=&-u\biggl(a_{i}\bigl(1+v^{2}\bigr)^{-1-\varepsilon /2}+b_{i}
\biggl[K_{\varepsilon }(v)-\bigl(1+v^{2}\bigr)^{-1-\varepsilon /2}\biggr]
\biggr)+O\bigl(u^{2}\bigr)\  \\
&=&-w\biggl(a_{i}\bigl(1-C_{\varepsilon }(v)\bigr)+b_{i}C_{\varepsilon }(v)
\biggr)+O\bigl(u^{2}\bigr)\ .
\end{eqnarray}
Here we have used the new coupling constant $w=uK_{\varepsilon }(v)$ and
the definition (\ref{Cfunkt}) of the crossover function $C_{\varepsilon }(v)$
. We mention that $w=u$ for $v=0$ and $w=\bar{u}/2$ for $v=\infty $, as well
as $C_{\varepsilon }(0)=0$ and $C_{\varepsilon }(\infty )=1$. $
C_{\varepsilon }(v)$ crosses over monotonically between this two values.
Especially in three dimensions it is a simple rational function of $v^{2}$:
\begin{equation}
C_{3}(v)=\frac{v^{2}\bigl(5+2v^{2}\bigr)}{\bigl(1+v^{2}\bigr)\bigl(3+2v^{2}
\bigr)}\ .
\end{equation}
The renormalization group equation~(\ref{RGG}),
\begin{equation}
\biggl[{\cal D}_{\mu }+\frac{N+\widetilde{N}}{2}\gamma \biggr]G_{N,
\widetilde{N}}=0 \ , \label{RGG2}
\end{equation}
follows as usually from the independence of the unrenormalized Green's
functions from the external mass scale $\mu $. Here we have to note that the
mass $m$ is by definition (\ref{mass}) a function of the bare
parameters only. Therefore, its Gell-Mann-Low function vanishes. Thus, we have the renormalization group differential operator
\begin{equation}
{\cal D}_{\mu }=\mu \partial _{\mu }+\beta _{w}\partial _{w}+\beta
_{v}\partial _{v} \ . \label{RGOp2}
\end{equation}
To one-loop order we obtain
\begin{eqnarray}
\beta _{w} &=&\biggl(-\varepsilon -C_{\varepsilon }(v)+\frac{
42-5C_{\varepsilon }(v)-C_{\varepsilon }(v)^{2}}{12}w\biggr)w  \label{betaw2}
\\
\beta _{v} &=&\biggl(-1+\frac{2+C_{\varepsilon }(v)}{12}w\biggr)v\ ,
\label{betav2}
\end{eqnarray}
and
\begin{equation}
\gamma =-\frac{2+C_{\varepsilon }(v)}{12}w \ .  \label{gam2}
\end{equation}

The flow equations
\begin{eqnarray}
l\frac{d}{dl}\bar{w}(l) &=&\beta _{w}\bigl(\bar{w}(l),\bar{v}(l)\bigr)\
,\qquad \bar{w}(1)=w\ ,  \label{wflow2} \\
l\frac{d}{dl}\bar{v}(l) &=&\beta _{v}\bigl(\bar{w}(l),\bar{v}(l)\bigr)\
,\qquad \bar{v}(1)=v\ ,  \label{vflow2}
\end{eqnarray}
and their solutions show of course the (unstable) isotropic and the (stable)
directed percolation fixed points for $v_{\ast }=0$ and $v_{\ast }=\infty$, 
respectively, and the continuous crossover between both. 

Upon defining the flowing amplitude function $X(l)$ by
\begin{equation}
l\frac{d}{dl}\ln X(l)=\gamma \bigl(\bar{w}(l),\bar{v}(l)\bigr)\ ,\qquad
X(1)=1\ ,  \label{Xflow}
\end{equation}
the scaling form of the Green's functions follows from the
renormalization group equation~(\ref{RGG2}) as
\begin{eqnarray}
G_{N,\widetilde{N}}(\{{\bf x}\},m;w,v;\mu ) &=&\Bigl(m^{d-2}X(m/\mu
)\Bigr)^{(N+\widetilde{N})/2}G_{N,\widetilde{N}}\bigl(\{m{\bf x}\};\bar{w}
(m/\mu ),\bar{v}(m/\mu )\bigr)  \nonumber \\
&\simeq &\Bigl(m^{d-2}\frac{X(m/\mu )}{2\bar{v}(m/\mu )}\Bigr)^{(N+
\widetilde{N})/2}\bar{G}_{N,\widetilde{N}}\bigl(\{m{\bf x}_{\bot },m^{2}\bar{
\lambda}(m/\mu )t\};\bar{w}(m/\mu )\bigr)\ ,  \label{G-Lsgn}
\end{eqnarray}
where the last form holds asymptotically for $\bar{v}(m/\mu )\gg 1$, and $
\lambda /\bar{\lambda}(l)=l\bar{v}(l)/v$.

We conclude this section with the determination of the parameter $m$ as a function of
the temperature $\tau $. For the renormalized temperature derivative we have
the renormalization group equation
\begin{equation}
\biggl[{\cal D}_{\mu }-\kappa \biggr]P=0  \label{P-Rg}
\end{equation}
with $\kappa =\gamma -\gamma _{\tau }$. Its solution is given by
\begin{equation}
P(m/\mu ;w,v)=Y(m/\mu )P(1;\bar{w}(m/\mu ),\bar{v}(m/\mu ))\ ,  \label{P-Lsg}
\end{equation}
where the amplitude function $Y$ is determined by the differential equation
\begin{equation}
l\frac{d}{dl}\ln Y(l)=-\kappa \bigl(\bar{w}(l),\bar{v}(l)\bigr)\ ,\qquad
Y(1)=1\ .  \label{Yflow}
\end{equation}
$P$ comprises the main (exponential) contribution of the crossover-scaling
of $\left. \partial \tau /\partial m^{2}\right| _{w,v,\mu }$. The amplitude $
P(1;w,v)$ is given to one-loop order by Eq.\ (\ref{Pfunkt}). The temperature
$\tau $ results then from the solution of the differential equation $\left.
\partial \tau /\partial m^{2}\right| _{w,v,\mu }=P(m/\mu ;w,v)$. The integration of all the flow-equations has to be done numerically and leads, qualitatively, back to the numerical results of FTS~\cite{FTS94}.

\section{Epilogue}

Using field theoretic methods, we have analyzed the connectivity behavior of
random resistor-diode networks near the percolation critical point. We found
that the introduction of positive and negative diodes oriented to a
privileged direction in space with unequal probabilities leads to a
crossover to the directed percolation problem, whereas a distribution of
diodes with equal probabilities results only in elongated isotropic
percolation clusters. In the latter case, a simple rescaling of the
privileged direction maps the problem to isotropic percolation. A slightly
different distribution of the diodes introduces a further relevant variable
with a new scaling dimension. We have calculated this scaling dimension to $
O(\varepsilon ^{2})$ in an $\varepsilon $-expansion around six
dimensions. An interpolation resulting from this $\varepsilon
$-expansion and an exact value at one dimension leads to a formula
that compares very well with a recent simulational  result in two
dimensions. It would be very interesting to perform simulations
also in higher dimensions in order to compare these with our result.

In Sec.~V we have reconsidered the theory by Frey, T\"{a}uber, and Schwabl for the crossover from isotropic to directed percolation. Some shortcomings are
corrected and it is demonstrated, how one can perform crossover
calculations consistently by using a type of extended minimal renormalization.

\acknowledgments 
This work has been supported in part by the SFB 237 ,,Unordnung und gro\ss e Fluktuationen`` of the Deutsche Forschungsgemeinschaft.

\appendix

\section*{Two-Loop Calculation}

In this appendix we present briefly the main part of the two-loop
calculation of the new renormalization constants $Z_{v}$ and $Y_{v\tau }$.
The dimensionaly regularized parameter integral
\begin{eqnarray}
I(a,b,c) &=&\int_{{\bf p,q}}\frac{1}{(a+{\bf p}^{2})(b+{\bf q}^{2})(c+({\bf 
p+q})^{2})}  \nonumber \\
&=&\frac{G_{\varepsilon }^{2}}{6\varepsilon }\biggl(\Bigl(\frac{1}{%
\varepsilon }+\frac{25}{12}\Bigr)\bigl(a^{3-\varepsilon }+b^{3-\varepsilon
}+c^{3-\varepsilon }\bigr)-3abc  \nonumber \\
&&-\Bigl(\frac{3}{\varepsilon }+\frac{21}{4}\Bigr)\bigl(a^{2-\varepsilon
}(b+c)+b^{2-\varepsilon }(a+c)+c^{2-\varepsilon }(b+c)\bigr)\biggr)\ ,
\label{ParInt}
\end{eqnarray}
introduced in Ref.~\cite{BrJa81}, plays a fundamental role in the calculation.
Its derivatives
\begin{equation}
I_{lmn}=\frac{(-1)^{l+m+n-3}}{(l-1)!(m-1)!(n-1)!}\left. \frac{\partial
^{l+m+n-3}I(a,b,c)}{\partial a^{l-1}\partial b^{m-1}\partial c^{n-1}}\right|
_{a=b=c=1}  \label{Ilmn}
\end{equation}
are extensively used in the following. Particularly simple are the integrals
\begin{equation}
I_{n}=\int_{{\bf p}}\frac{1}{(1+{\bf p}^{2})^{n}}=\frac{8(-1)^{n-1}G_{
\varepsilon }}{(n-1)!(4-\varepsilon )(2-\varepsilon )\varepsilon }\left.
\frac{\partial ^{n-1}a^{2-\varepsilon /2}}{\partial a^{n-1}}\right| _{a=0}\ .
\label{In}
\end{equation}

We start with the self-energy diagram displayed in Fig.~2(a). Its value is given by
\begin{eqnarray}
\Sigma ^{(2a)}({\bf q}) &=&-\frac{g^{4}}{2}\int_{{\bf k,p}}G({\bf q-k})G(
{\bf k})^{2}G({\bf p})G({\bf k-p})  \nonumber \\
&=&-\frac{g^{4}}{2}\int_{{\bf k,p}}\frac{1}{[\bar{\tau}+({\bf \bar{q}-k}
)^{2}][\bar{\tau}+{\bf k}^{2}]^{2}[\bar{\tau}+({\bf p+}i{\bf v})^{2}][\bar{
\tau}+({\bf k-p})^{2}]}\ ,  \label{SE2a}
\end{eqnarray}
where we have shifted ${\bf k\rightarrow k-}i{\bf v}$, and
defined $\bar{\tau}=\tau +{\bf v}^{2}$ and ${\bf \bar{q}}={\bf q}+2i{\bf v}$. We expand the last integral in ${\bf q}$ and ${\bf v}$ to second order and
find
\begin{eqnarray}
\Sigma ^{(2a)}({\bf q}) &=&-\frac{g^{4}}{2}\biggl(I_{113}+\Bigl(\frac{d-4}{d}
I_{123}+\frac{4\bar{\tau}}{d}I_{133}\Bigr){\bf v}^{2}  \nonumber \\
&&+2\Bigl(I_{2}I_{4}+\bar{\tau}I_{124}-I_{114}-I_{123}\Bigr)\frac{i{\bf 
v\cdot \bar{q}}}{d}-\Bigl(\frac{d-4}{d}I_{114}+\frac{4\bar{\tau}}{d}
I_{115}\Bigr){\bf \bar{q}}^{2}\biggr)\ .  \label{SE2a2}
\end{eqnarray}
In the same fashion we get for the diagram shown in Fig.~2(b)
\begin{eqnarray}
\Sigma ^{(2b)}({\bf q}) &=&-g^{4}\int_{{\bf k,p}}G({\bf q-k})G({\bf q-p})G(
{\bf k})G({\bf p})G({\bf k-p})  \nonumber \\
&=&-g^{4}\tau ^{-\varepsilon }\biggl(I_{122}+\Bigl(\frac{d-4}{d}I_{222}+
\frac{4\bar{\tau}}{d}I_{223}\Bigr){\bf v}^{2}  \nonumber \\
&&-2\Bigl(\frac{d-6}{d}I_{123}+\frac{4\bar{\tau}}{d}I_{124}+\frac{1}{d}
I_{3}^{2}+\frac{\bar{\tau}}{d}I_{133}\Bigr){\bf \bar{q}}^{2}\biggr)\ .
\label{SE2b}
\end{eqnarray}
Using Eqs.~(\ref{Ilmn},\ref{In}) we obtain from Eqs.~(\ref{SE2a2},\ref{SE2b}) the singular contributions to the two-loop vertex function after renormalization according to the scheme~(\ref{Ren1a},\ref{Ren1b},\ref{Ren1c}) as
\begin{eqnarray}
\Gamma _{1,1}^{(2-loop)} &=&\frac{u^{2}}{\varepsilon ^{2}}\tau
^{-\varepsilon }\biggl[\Bigl(\frac{9}{4}+\frac{45\varepsilon }{16}\Bigr)\mu
^{2}\tau +\Bigl(\frac{5}{2}+\frac{25\varepsilon }{24}\Bigr)\frac{(\mu {\bf v)
}^{2}}{3}  \nonumber \\
&&+\Bigl(\frac{11}{6}+\frac{7\varepsilon }{72}\Bigr)\frac{{\bf q}^{2}}{6}
+\Bigl(\frac{23}{12}+\frac{13\varepsilon }{144}\Bigr)\frac{2i\mu {\bf v\cdot
q}}{3}\biggr]\ .  \label{2loop}
\end{eqnarray}

Using the renormalization constants to $O(u)$, Eqs.~(\ref{Z}-\ref{Zu},\ref
{Zv},\ref{Yvt}), we get from Eq.~(\ref{SelfEn2}), after the renormalization
$\mathaccent"7017{\Gamma }_{1,1}\rightarrow \Gamma _{1,1}=Z\mathaccent"7017{
\Gamma }_{1,1}$, the one-loop self-energy to $O(u^{2})$ as
\begin{eqnarray}
\Gamma _{1,1}^{(1-loop)} &=&-\frac{u}{\varepsilon }\tau ^{-\varepsilon /2}
\biggl[\biggl(1+\Bigl(\frac{9}{2}+\frac{11\varepsilon }{6}\Bigr)\frac{u}{
\varepsilon }\biggr)\mu ^{2}\tau +\biggl(1+\Bigl(5-\frac{5\varepsilon }{12}
\Bigr)\frac{u}{\varepsilon }\biggr)\frac{(\mu {\bf v)}^{2}}{3}  \nonumber \\
&&+\biggl(1+\Bigl(\frac{11}{3}-\frac{5\varepsilon }{12}\Bigr)\frac{u}{
\varepsilon }\biggr)\frac{{\bf q}^{2}}{6}+\biggl(1+\Bigl(\frac{23}{6}-\frac{
5\varepsilon }{12}\Bigr)\frac{u}{\varepsilon }\biggr)\frac{2i\mu {\bf v\cdot
q}}{3}\biggr]\ .
\end{eqnarray}

By adding $\Gamma _{1,1}^{(1-loop)}$, $\Gamma _{1,1}^{(2-loop)}$ and the renormalized zero-loop part, we get
\begin{equation}
\Gamma _{1,1}=\mu ^{2}\Bigl(Z_{\tau }\tau +Y_{v\tau }v^{2}\Bigr)+2iZ_{v}\mu
{\bf v}\cdot {\bf q+}Zq^{2}+\Gamma _{1,1}^{(1-loop)}+\Gamma
_{1,1}^{(2-loop)}+O(\varepsilon ^{0},u^{3})\ .
\end{equation}
We see that the nonprimitive divergencies $\sim \ln \tau $ of $\Gamma
_{1,1}^{(1-loop)}$ and $\Gamma _{1,1}^{(2-loop)}$ cancel (as a check of a
correct calculation) and find finally the renormalization constants cited in
Eqs.~(\ref{Z},\ref{Zt},\ref{Zv},\ref{Yvt}).


\begin{figure}[h]
\begin{center}
\epsfig{file=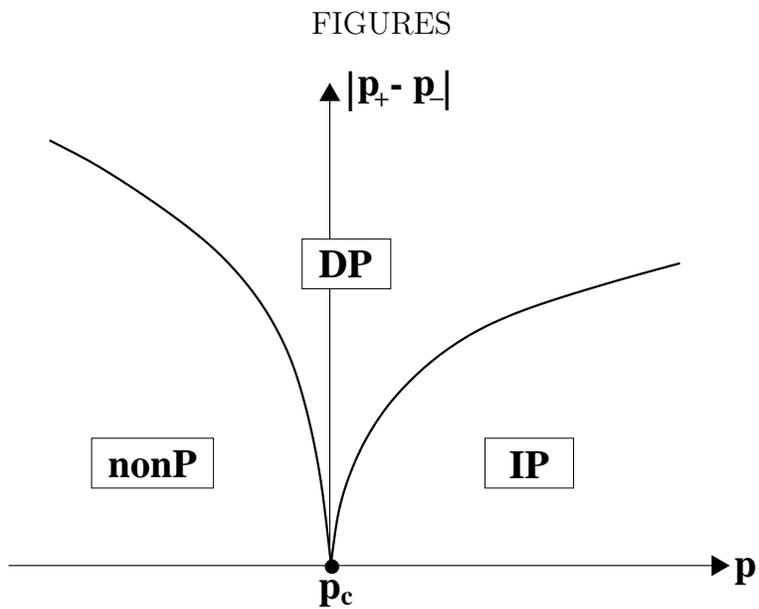,width=10cm}
\end{center}
\caption{\label{phaseDiag} Generic phase diagram for $p_+ + p_- = \mbox{const} \leq 1-p$. The nomenclature is the following, $p_c$: critical point, $nonP$: non-percolating phase,
$DP$: phase of directed percolation, $IP$: phase of (elongated) isotropic percolation.}
\end{figure}
\clearpage
\begin{figure}[h]
\begin{center}
\epsfig{file=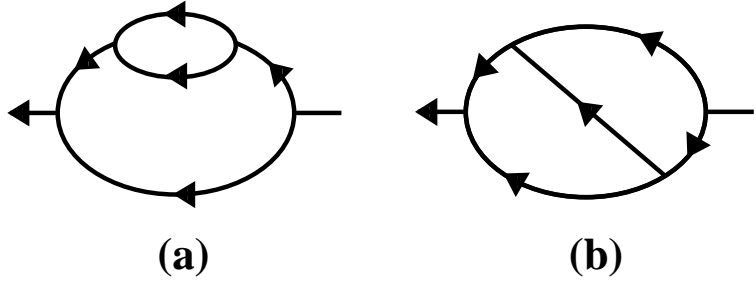,width=10cm}
\caption{\label{diagramsPic} Two-loop self-energy diagrams.}
\end{center}
\end{figure}

\end{document}